\documentclass{article}

\usepackage{arxiv}

\usepackage[utf8]{inputenc} 
\usepackage[T1]{fontenc}    
\usepackage{hyperref}       
\usepackage{url}            
\usepackage{booktabs}       
\usepackage{amsfonts}       
\usepackage{nicefrac}       
\usepackage{microtype}      
\usepackage{lipsum}
\usepackage{graphicx}
\graphicspath{ {./images/} }

\title{Morphology and structural properties of thin rubrene crystallites grown on graphite}

\author{
 Moha Naeimi \textsuperscript{a, b}, Katharina Engster \textsuperscript{a, b}, Ingo Barke \textsuperscript{a, b}, Sylvia Speller \textsuperscript{a, b} \\
 a Institute of Physics, University of Rostock, Albert-Einstein-Str. 23, 18059 Rostock,
Germany.\\
 b Department “Life, Light and Matter”, University of Rostock, Albert-Einstein-Str. 25,
18059 Rostock, Germany.\\
  \texttt{sylvia.speller@uni.rostock.de} \\
}

\begin{document}
\maketitle
\begin{abstract}
Crystallization of rubrene, progressing from an amorphous phase to a triclinic meta-stable and ultimately to the orthorhombic stable phase, offers broad applications not only in organic electronic devices but also for in-depth studies of optical and electronic properties, including exciton distribution and dynamics. We investigate the crystallization of rubrene on highly oriented pyrolytic graphite (HOPG), aiming at the growth of the preferred rubrene orthorhombic phase, which has been reported to have one of the highest charge mobilities in organic semiconductors. This is achieved through controlled heating and enhanced partial pressure. Through precise control of the initial deposition on the substrate, we investigate the growth habit of rubrene crystals by high-rate heat treatment beyond the second crystallization temperature. Furthermore, this work addresses thermal stability and photodegradation across various morphologies.
\end{abstract}


\section{Introduction}

Rubrene crystalline structures have been subject of intense research in the context of organic electronics \cite{McGarry2013}, e.g. for field-effect transistors \cite{Kim2007, Duan2020} and organic light emitting diodes \cite{Wang2023, Wang2010}. Rubrene is a small, organic molecule which can be thought to be based on tetracene with 4 additional phenyl rings. It shows various interesting electronic properties such as high charge mobility \cite{vanderLee2022, Zhang2010, Nitta2019}, efficient singlet fission \cite{Wu2021, Breen2017} with long life time and long diffusion length \cite{Finton2019}, triplet-triplet annihilation and fusion \cite{Baronas2022, Engmann2019} and triplet energy transport \cite{Baronas2022}. Due to these properties, rubrene is a promising candidate for energy and exciton transfer both in lateral and vertical direction with respect to the surface. Rubrene singlet fission has been widely studied in the past decades. It is well known that singlet fission in rubrene is more efficient in crystalline stable phases \cite{Ma2012} while in amorphous structures it is significantly lower \cite{Chen2020}. Hence, crystalline structures are necessary to provide a playground to extract its unique electronic properties.

Growth of rubrene crystalline domains both in platelets and spherulites \cite{Fusella_2017, Horike2014} has been the subject of several studies aiming for large lateral dimensions by tailored heating treatment. However, preparing extended rubrene crystals directly on a substrate is not simple because the structures have the tendency to become trapped in a metastable state of low order or in an unfavorable crystal structure during the preparation process \cite{Su2012}. The challenge essentially lies in the fact that providing sufficient thermal energy to overcome this metastable state easily results in massive desorption, eventually evaporating all of the available material on the substrate. To overcome this, a variety of methods have been proposed, which are based on temperature treatments or epitaxial methods \cite{Chang2015, AkinKara2022, Luo2007, Park2007, Kratzer2019}. A key  idea is to reach the crystallization temperature while preventing desorption, by providing a high rate heat transfer to amorphous rubrene in the local environment of the substrate.

Most of the studies use silicon substrates with a native oxide layer and different adlayers. The aim of this work is to control the structure and morphology of highly crystalline, thin and flat rubrene crystallites on bare HOPG. In particular, well separated crystals with laterally extended single domains are sought. Special focus is on simplicity, reproducibility, purity and flexibility. Such samples shall be suitable for exciton transfer studies using surface science techniques such as local photoelectron spectroscopy \cite{Nakayama2012}. Important is to prevent optical waveguiding \cite{Ma2021}, i.e. along at least one dimension crystal should measure only a few 10 nm.

Among the crystal phases of rubrene, the well-known orthorhombic phase had been reported to have a very high charge mobility of 43 $cm\textsuperscript{2}/Vs$ \cite{Saeki2008}, one of the highest charge mobilities reported in organic semiconductors. Rubrene molecules are arranged in this phase in a way such that their intermolecular interactions are minimized by their phenyl rings. Although the orthorhombic crystal phase of rubrene is known to be the thermodynamically stable phase, in contrast to the triclinic phase which is metastable, it has been a challenge to grow in thin, yet ultra-smooth crystals directly on a surface.

In the present paper, we investigate the control of the growth of rubrene crystalline structures on highly  oriented pyrolytic graphite (HOPG) using the method of abrupt-heating \cite{Lee2011}. We use HOPG as a substrate, because it has a well-defined, flat surface on the atomic scale, it is conductive, and it is inert which simplifies sample handling. Apart from rubrene, no other co-materials such as buffer layers are used throughout this work to aid growth \cite{Verreet2013, Tan2023, Fusella_2017, Foggiatto2019}. We also address thermal stability and occurrence of photo-oxidation processes.

\begin{figure}
    \centering
    \includegraphics[width=450pt]{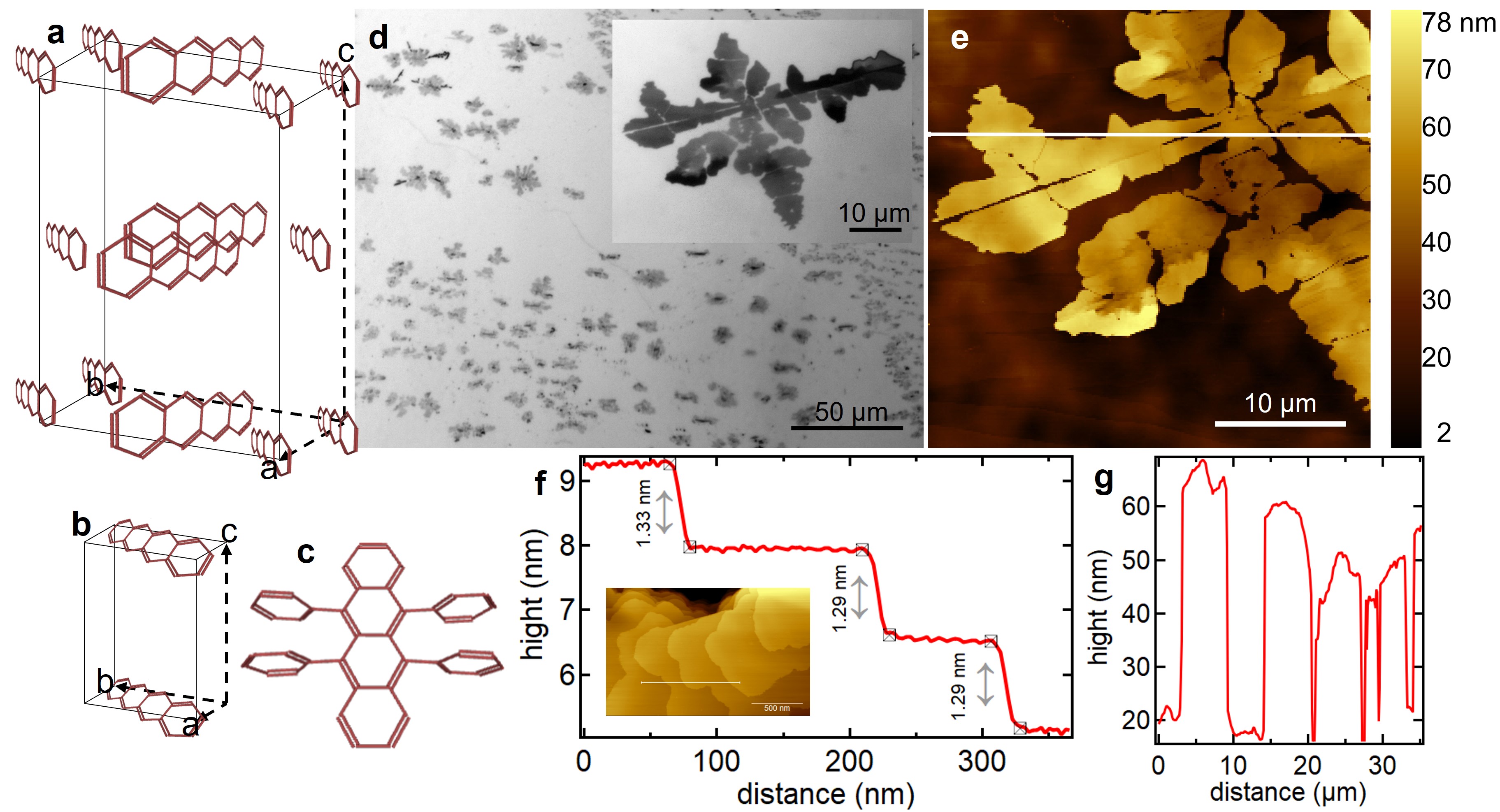}
    \caption{(a - c) Schematic representations of rubrene orthorhombic and triclinic unit cell along with rubrene molecule, respectively. In the stacking, only the tetracene backbone is shown for simplicity. (d) Bright-field image with a magnified image of a single crystal prepared by spin-coating and high-rate heating (e) AFM image of the crystal in (d). (f) AFM image showing step heights of $\approx$1.3 nm (g) Line profile at the location indicated in (e).}
    \label{fig:spin_AFM_Unitcell}
\end{figure}

\section{Experimental section}

Rubrene powder with a purity of $99.999,/\%$ (Sigma-Aldrich) was used as received and the container was kept in a desiccator to prevent oxidation. For preparation using the high-rate heating approach, also referred to as abrupt heating, the sample is directly covered with rubrene, either using spin-coating or high-vacuum evaporation (see below for details). Then the sample was placed onto a pre-heated hot plate. The temperature of the sample was carefully controlled and verified by a K-type thermocouple to anneal the amorphous phase at the second crystallization stage of 170$^{\circ}$C for different heating times from 1 min to 1 hour. For spin-coating, we used toluene and chloroform as base solvents with different solute concentrations ranging from $10^{-3}$ to $10^{-5}$ molar. For high-vacuum evaporation, the rubrene powder was placed inside a heated glass vessel, whose target temperature was changed between 170 to 200$^{\circ}$C to control the evaporation rate. The substrate was kept at room temperature. The base pressure of the evaporation chamber was of the order of $10^{-7}$ mbars. HOPG was cleaved in air with adhesive tape and no further annealing was done. 

After preparation, the crystals were investigated by optical microscopy (Zeiss Axio scope.A1), fluorescence microscopy (Olympus IX73 inverted fluorescence microscope), polarization optical microscopy (POM) (Zeiss Axiolab.A5) and atomic force microscopy (AFM)(Park Systems NX20 and XE100).

For photoemission electron microscopy (PEEM) we introduced the samples to a PEEM (Focus IS-PEEM) with the base pressure of $10^{-10}$ mbar and used the 3rd harmonics of a tunable Ti:Sa femtosecond laser (Mira 900F) yielding photons with energies equal to 4.6 eV (266 nm). The light angle of incident was 67 degrees.

Visualisation and analysis was done by Gwyddion \cite{Necas2012}and Igor Pro (Wavemetrics). The evaluation of the area coverage was done by estimation of the apparently covered area, marked manually, using ImageJ software.

\section{Results and discussion}

\subsection{Spin coating and platelets}

Among the three different crystal phases of rubrene, there is a competition between triclinic and orthorhombic phases during heating. The monoclinic phase has the weakest intermolecular binding, and is not regarded in the present work \cite{Liu2023} since the growth does not happen on surfaces but only in solution \cite{Huang2010}. The triclinic phase is a result of low-rate heat transfer or heating at the first crystallization temperature (140$^\circ$C) \cite{Fielitz2016}, while a high-rate heat transfer to amorphous rubrene or heating at the second crystallization temperature (170$^{\circ}$C) yields the formation of the orthorhombic phase. Schemes of the rubrene molecule, the triclinic, and the orthorhombic structure are depicted in figure \ref{fig:spin_AFM_Unitcell}.a to c, along with the respective main crystal axes. Figure \ref{fig:spin_AFM_Unitcell}.d shows an overview brightfield image of the well separated crystals prepared by spin-coating 5 $\mu$L of a 1 mMol rubrene solution onto HOPG, using toluene as a solvent. Subsequently, the sample underwent a high-rate heating step for one minute on a preheated hot plate at a temperature of 170$^{\circ}$C, within a nitrogen-filled container. The inset of Figure \ref{fig:spin_AFM_Unitcell}.d is the magnified brightfield image of one of these crystals. As shown in Figure \ref{fig:spin_AFM_Unitcell}.e, which is an AFM image of the very same crystal in figure \ref{fig:spin_AFM_Unitcell}.d, the leaf-shape platelets are quite thin, their height typically measures a few 10 nm. The line profile in \ref{fig:spin_AFM_Unitcell}.e is indicated in Figure \ref{fig:spin_AFM_Unitcell}.f. Figure \ref{fig:spin_AFM_Unitcell}.f shows a high resolution AFM image and a line profile at the top of the crystal revealing single-layer steps. Step heights are about $\approx$1.3 nm, implying an orthorhombic phase with the c-axis of the unit cell oriented perpendicular to the surface \cite{ElHelou2010}.

\begin{figure}
    \centering
    \includegraphics[width=400pt]{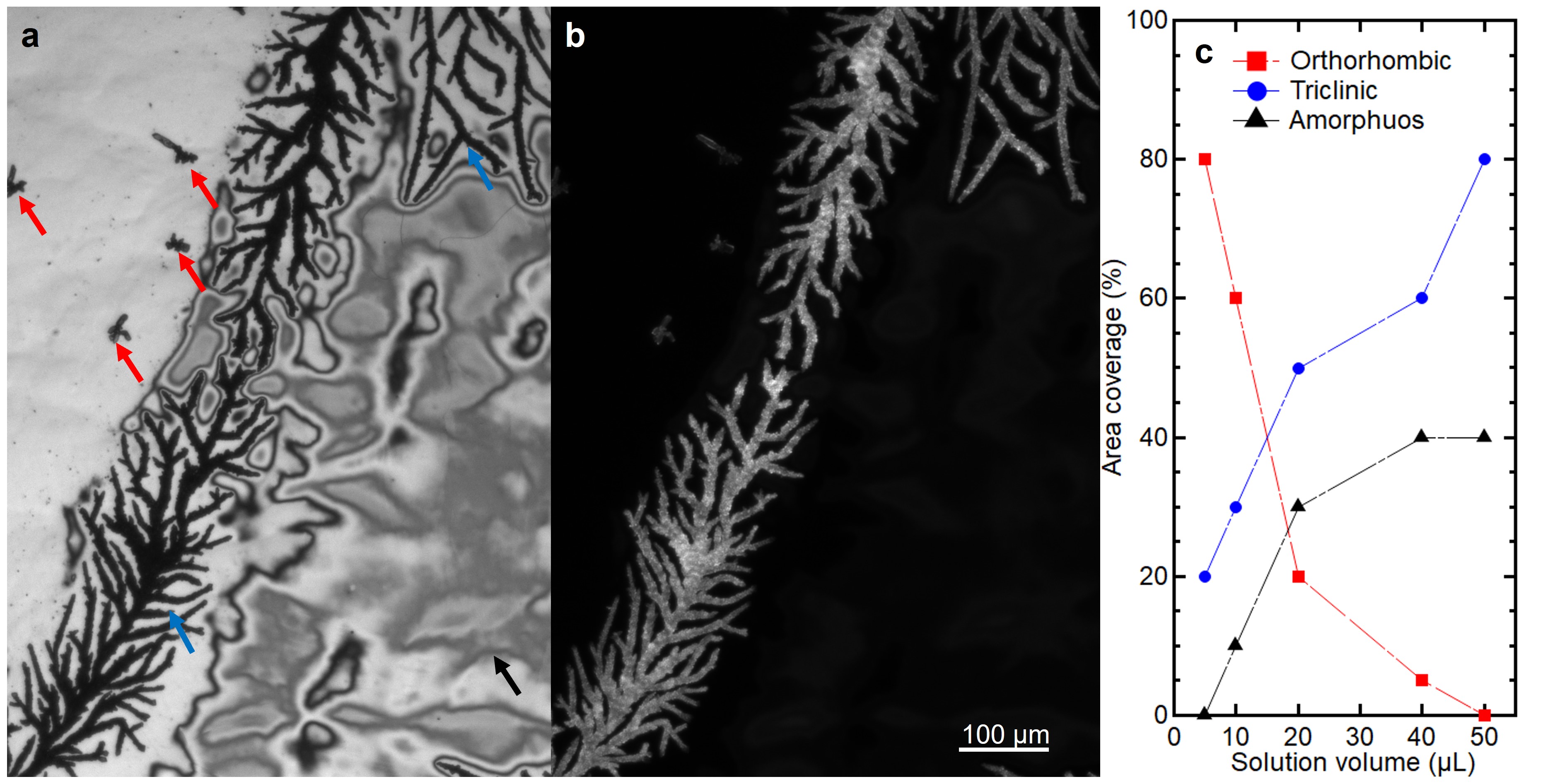}
    \caption{Sample prepared by spin-coating and high-rate heating. (a and b) Bright-field and fluorescence images showing three different morphologies of rubrene formed on HOPG. The triclinic domains (marked blue) show high fluorescence intensity and form in areas with high rubrene coverage in contrast to orthorhombic single crystallites (marked red) which grow in areas with low rubrene coverage. (c) A semi-quantitative analysis of rubrene polymorphs formed after high-rate heating versus solution coverage spin-coated on HOPG.}
    \label{fig:spin_cov_bf_flur}
\end{figure}

A competition between single- and polycrystalline orthorhombic, and the polycrystalline triclinic phase was observed, which can be controlled by increasing the volume of spin-coated solution. Figure \ref{fig:spin_cov_bf_flur}.a and \ref{fig:spin_cov_bf_flur}.b show overview brightfield and fluorescence images of an exemplary region with local high (bottom right) and low (top left) coverage, respectively, prepared by spin coating of 20 uL rubrene-toluene solution, subsequently subjected to high-rate heating for one minute at 170$^{\circ}$C. It shows coexisting  orthorhombic, triclinic and amorphous domains. As indicated in figure \ref{fig:spin_cov_bf_flur}.c, spin coating of more solution onto HOPG favours the polycrystalline triclinic phase, while orthorhombic single crystallites are the result of low solution coverage. A reasoning why more solution, i.e., more molecules, reduce the emergence of the orthorhombic phase, can be done considering the balance between flux or delivery rate and diffusion. In spin-coating the concentration and volume of the molecules and solution determines the rate of molecules delivered. If the delivery is fast, many nucleation centers occur at high density in a short time period. Upon further growth of these nuclei domain boundaries will result because the different islands / grains cannot be monolithically connected. This favors phases with lower order, such as the triclinic polymorph.

\begin{figure}
    \centering
    \includegraphics[width=350pt]{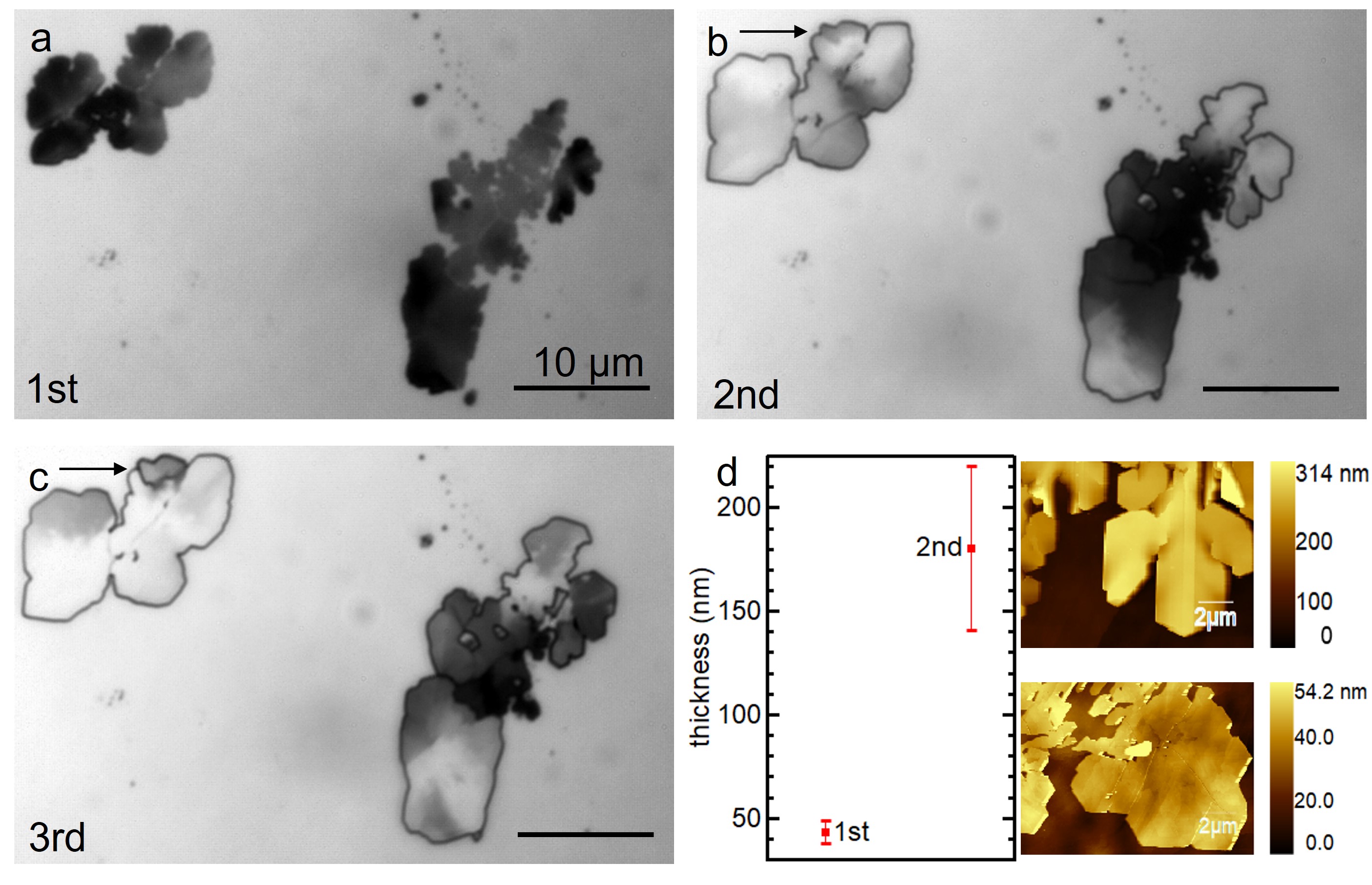}
    \caption{Sample prepared by spin-coating and consecutive high-rate heating. (a) - (c) A series of bright-field images showing the growth evolution through consecutive heating (d) Average growth in thickness from different samples with AFM images of a sample before and after 2nd high-rate heating. The error bars indicate the standard deviation of the thickness of a selection of $>$~20 measured samples for each heating stage.}
    \label{fig:cons_heat_bf_AFM}
\end{figure}

One factor that determines the thickness of crystals is the total number of high-rate heating steps. Figure \ref{fig:cons_heat_bf_AFM} shows a series of brightfield images of a crystal prepared by spin coating of 0.5 uL rubrene-toluene solution on HOPG and treated consecutively with high-rate heating, each for 60 seconds at 170$^{\circ}$C. After the first heat treatment, the crystal thickness, based on brightfield-AFM correlations, is estimated to be 40 nm, while the second heat treatment changed the crystal thickness to a couple of hundred nm while the lateral size slightly increased. A third high-rate heat treatment was performed and resulted a sub-10 nm thickness change and for the fourth heating treatment, no change in thickness was observed anymore. The final platelets show step heights that match the orthorhombic crystalline phase of rubrene. Figure \ref{fig:cons_heat_bf_AFM}.d shows the average growth of the single crystals by measuring the thickness of various samples with AFM, through the first and second high-rate heating step. We do not observe indications of oxidation for these crystals (see below).

To summarize this section, we obtain well-separated single orthorhombic platelets with exceptionally flat surfaces when starting from a solution which is deposited on HOPG through spin coating with low coverage, followed by high-rate heat treatment. The same preparation conducted with elevated coverage yields an increased fraction of triclinic crystals. Interestingly, the same preparation on silicon with a native oxide layer results in amorphous, droplet-like structures, with occasional triclinic polycrystalline phases, independent of the coverage. We attribute this observation to the fact that HOPG has an atomically well-defined surface, while the oxide layer on silicon is amorphous, hence impeding the initial ordered growth of large orthorhombic domains.

\subsection{Photo-oxidation assessment}
\begin{figure} [ht]
    \centering
    \includegraphics[width=300pt]{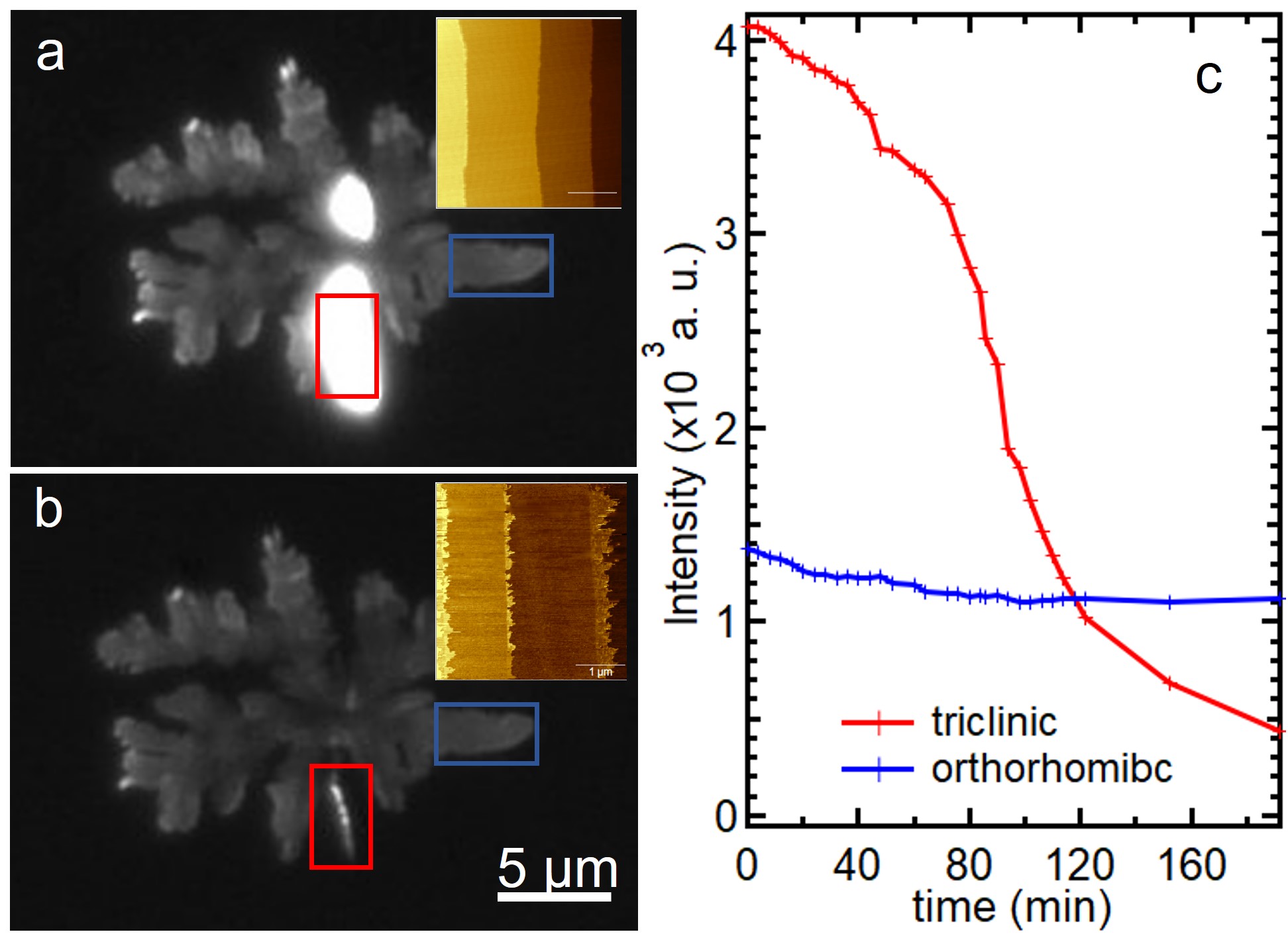}
    \caption{Photo-bleaching of rubrene polycrystalline structure prepared by spin-coating and high-rate heating. (a and b) Fluorescence image of the rubrene polycrystal before and after light exposure. The insets are high resolution AFM images of the orthorhombic part (marked blue) of the polycrystal showing the photo-oxidation occurring at the step edges. For the triclinic part, no change of morphology is visible by AFM.  d: Fluorescence degradation profile of two different crystal phases of rubrene during 3 hours of light exposure (excitation wavelength: ~ 540 nm).}
    \label{fig:photo_deg_flur_AFM}
\end{figure}

The fluorescence intensity of rubrene crystalline phases strongly depends on their structure and orientation \cite{Irkhin2012}. We introduced a large number of our crystals into a fluorescence microscope and compared their intensity and monitored their photo bleaching behavior. As we observed, the fluorescence intensity is significantly higher for the triclinic phase compared to the flat platelets belonging to the orthorhombic phase, which can be used for quick and easy assignment for crystalline phases (along with supporting AFM investigations). Since the triclinic phase of rubrene has a polycrystalline nature by itself \cite{Euvrard2022} and the orthorhombic phase usually tends to grow in a single crystalline domain, it is interesting to compare the photo-degradation behaviour between those two. Figure \ref{fig:photo_deg_flur_AFM}.a and \ref{fig:photo_deg_flur_AFM}.b show fluorescence images of a polycrystalline rubrene assembly prepared by spin coating 5 uL of rubrene-toluene solution onto HOPG, followed by high-rate heating at 170$^{\circ}$C for one minute. One can see that the triclinic part (red marked) yields high fluorescence intensity (\ref{fig:photo_deg_flur_AFM}.a) but after three hours of light exposure (\ref{fig:photo_deg_flur_AFM}.b), it degrades to almost zero. Figure \ref{fig:photo_deg_flur_AFM}.d shows the intensity of the marked parts versus exposure time. After two hours of illumination, the orthorhombic part seems to stay stable with no sign of further degradation. Interestingly, the triclinic phase shows an inflection point, meaning the degradation of fluorescence intensity is most rapid between one and two hours after start of exposure by the microscope (100 Watt Halogen lamp).

The difference in rate of photo-degradation between these two phases could be explained by the defect density \cite{Kytka2007, Mastrogiovanni2014} which is much larger for triclinic structures. Insets of figures \ref{fig:photo_deg_flur_AFM}.a and \ref{fig:photo_deg_flur_AFM}.b show the AFM height image of a related orthorhombic crystal indicating that photo-oxidation predominantly occurs at the step edges and defect spots on the surface. Previous reports have indicated that this process can also penetrate into the bulk \cite{Kytka2007, Mastrogiovanni2014, Fumagalli2011}. The AFM phase image is also suitable to demonstrate this oxidation behaviour, because the AFM tip experiences variations in adhesion and friction forces for different chemical compositions \cite{Werner2023}. These differences in interaction result in phase shifts that could be observed in the phase image in figure S1.  

The transition dipole moment of rubrene is along the short flat axis of the molecule (i.e. horizontally within the plane of figure \ref{fig:spin_AFM_Unitcell}a). Due to the polycrystalline nature of the triclinic phase of rubrene, the transition dipole moments of the molecules have random orientations and the chance of matching these and the polarization of the light (which is in-plane in our case) is significantly higher than for the orthorhombic phase whose dipole moment is oriented perpendicular to the polarization of our excitation light. Also, distinct polymorphs have much larger effect in luminescence yield  compared to correlation between thickness and luminescence. In contrast, in orthorhombic single crystals, the crystal thickness and luminescence are found to be correlated (Figure S2).

\begin{figure}
    \centering
    \includegraphics[width=450pt]{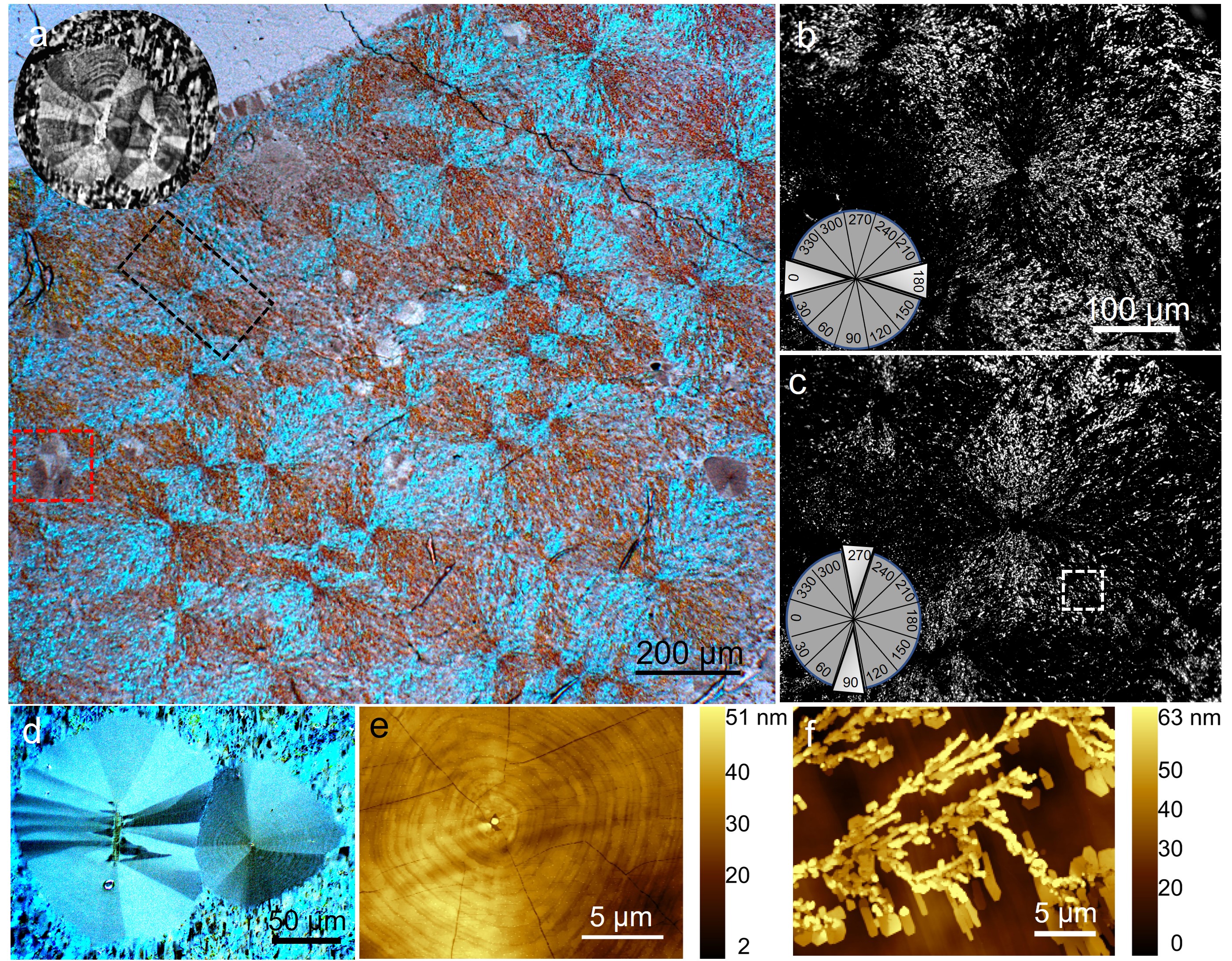}
    \caption{a: Polarization optical microscope image of rubrene crystal colonies grown with almost radial symmetry. The orthorhombic platelets are embedded in a bath of needle-shape crystals forming triclinic spherulites. The inset of image a is a PEEM image of the marked platelet showing different photoemission yield for different crystal domains. b and c: Fluorescence images of the triclinic spherulites. The polarization of the emitted light is filtered along 0 (horizontally) and 90 (vertically), respectively. d: POM image of the marked platelet in a. The sample is rotated 45 degrees clockwise with the same polarization orientation in a. e: AFM image of the same platelet in d. f: AFM image of the triclinic needle-shape crystal colonies marked in c with dashed square.}
    \label{fig:sph_plat_flur_PEEM_AFM}
\end{figure}

\subsection{Vacuum deposition and spherulites}
The initial rubrene coverage required for high-rate heating preparation can be provided by various ways. Apart from spin-coating, as described above, vacuum evaporation of a thin layer of amorphous rubrene on HOPG is another option. This results in structures summarized in figure\ref{fig:sph_plat_flur_PEEM_AFM} showing polarization optical microscopy, fluorescence, AFM, and PEEM images. Large and relatively rough crystalline domains occur, where the two most dominant structures are spherulite-like crystal colonies (figure \ref{fig:sph_plat_flur_PEEM_AFM}.a) and platelets (figure \ref{fig:sph_plat_flur_PEEM_AFM}.d and \ref{fig:sph_plat_flur_PEEM_AFM}.e). Figure \ref{fig:sph_plat_flur_PEEM_AFM}.a is a polarization optical microscope image of rubrene orthorhombic platelets embedded in needle-shape crystal colonies prepared by evaporating a 20 nm layer of rubrene on HOPG in high vacuum and subsequently high-rate heat treated in complete darkness. These structures which are known as triclinic spherulites \cite{Euvrard2022} are colonies of single crystals formed around the center of nucleation in an almost radially symmetric way. The inset of image \ref{fig:sph_plat_flur_PEEM_AFM}.a shows the PEEM image of the platelet (marked red) revealing, that the domains of the platelets have different photoemission yields, most likely caused by different absorption efficiencies of the laser light. As shown in figure \ref{fig:sph_plat_flur_PEEM_AFM}.b and c, by rotating the orientation of the emission polarizer of the fluorescence microscope, only crystals with matching fluorescence polarization appear bright.

\begin{figure} [ht]
    \centering
    \includegraphics[width=350pt]{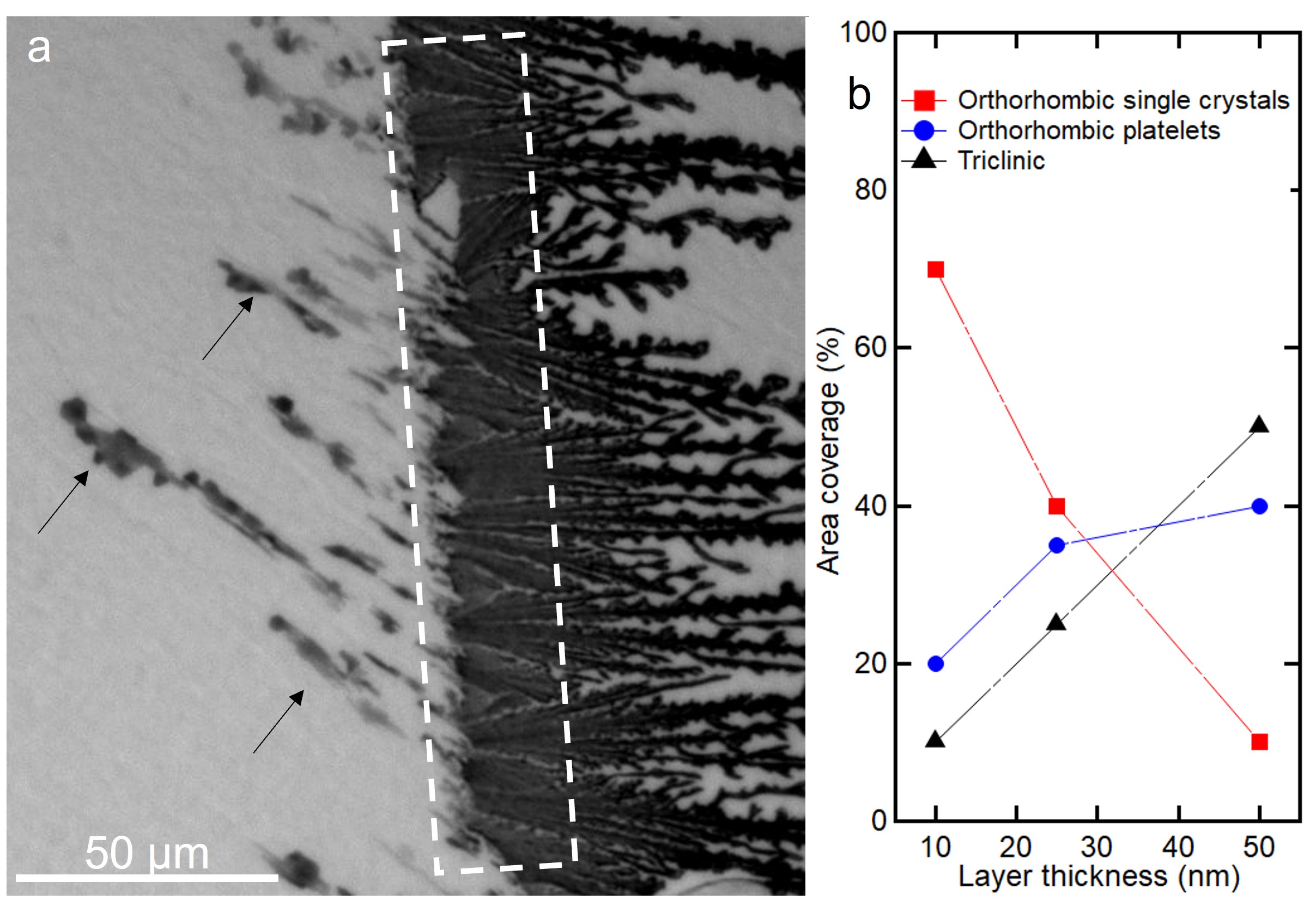}
    \caption{(a) Bright-field image of the area of a large rubrene coverage gradient, created using a shadow mask (left region covered, right region exposed). The growth of ultra-thin (thickness $<$10 nm) and ultra-flat (RMS roughness less than 100 pm) single crystals in the shadowed area is a result of surface diffusion during the evaporation phase. (b) Coverage of different possible structures versus the evaporated layer thickness on various samples.}
    \label{fig:orth_tric_transi_bf}
\end{figure}

We analyse the spherulites by means of polarization optical microscopy (in reflection, with entrance and exit polarizers 90° crossed) and fluorescence microscopy (in reflection, with only the luminescence light passing a polarizer). The imaging contrast with POM (figure \ref{fig:sph_plat_flur_PEEM_AFM}.a) is clearly along radial axes of the spherulites. Since the polarizers were crossed, the detected light will be dominated by structures which turn the electric field of the light by 90°, which is unlikely in absence of luminescence or birefringence. If the sample is turned 90° contrast swaps. Since the excitation light is from a LED with largely white light, we tend to attribute the contrast to a mixture of both. The imaging contrast of the fluorescence images (figures \ref{fig:sph_plat_flur_PEEM_AFM}.b and \ref{fig:sph_plat_flur_PEEM_AFM}.c) is however exclusively from the emitted, i.e., luminescent light. The illumination light has no preferred polarization. It reflects the radial symmetry of the spherulites containing small grains with triclinic structure, as has been found by x-ray diffraction [\cite{Euvrard2022}].

Figure \ref{fig:sph_plat_flur_PEEM_AFM}.d is the magnified POM image of the platelet marked in \ref{fig:sph_plat_flur_PEEM_AFM}.a. The sample is rotated 45 degrees clockwise with the same polarization orientation in \ref{fig:sph_plat_flur_PEEM_AFM}.a which results in changing the color and brightness of different domains. The AFM image of the platelet in figure \ref{fig:sph_plat_flur_PEEM_AFM}.e visualizes the growth mode from a center of nucleation to the edges of platelet with a relatively segmented rim. This is in contrast to platelets prepared by spin coating which yield ultra flat and smooth surfaces (compare figure \ref{fig:spin_AFM_Unitcell}.b and \ref{fig:sph_plat_flur_PEEM_AFM}.f). Figure \ref{fig:sph_plat_flur_PEEM_AFM}.f is an AFM image of the edge of a triclinic spherulite, revealing that these spherulites are consisting of small needle-shaped crystals having almost 100 nm of thickness. The step heights of both structures are $\approx$1.3 nm and reveal their orthorhombic nature.

As an intermediate conclusion, evaporation with subsequent high-rate heat treatment yields vastly different morphologies, and hence fluorescence properties, compared to solution coating as the first step. Evaporated rubrene results in a wide range of crystal morphologies, including the orthorhombic large crystalline domains known as spherulites and platelets that were repeatedly reported previously \cite{Verreet2013, Tan2023, Fusella2017, Fusella_2017, Foggiatto2019} using more complex preparation schemes. Often, considerable efforts have been spent to grow these domains such that they interconnect and fully cover the surface, e.g., by adding underlayers such as TPTPA and TPD \cite{Fusella2017}, PMMA \cite{Fusella_2017} and ITO \cite{Tan2023}. On bare HOPG, these orthorhombic platelets (figure \ref{fig:sph_plat_flur_PEEM_AFM}.d) are rather embedded in a bath of needle shape crystal colonies as large triclinic spherulite domains (figures \ref{fig:sph_plat_flur_PEEM_AFM}.a,  \ref{fig:sph_plat_flur_PEEM_AFM}.b and \ref{fig:sph_plat_flur_PEEM_AFM}.c). The orthorhombic platelets appear with different intensity and color in POM images due to their crystal order and orientation\cite{Lee2011}. Since in the orthorhombic phase the transition dipole moment is perpendicular to the surface, its fluorescence is completely suppressed which we indeed observe for the platelets (figure S3), such as those in figure \ref{fig:sph_plat_flur_PEEM_AFM}.d, at the same time revealing their perfect crystal ordering with the c axis perpendicular to the surface. \cite{Irkhin2012}.

\begin{figure}
    \centering
    \includegraphics[width=300pt]{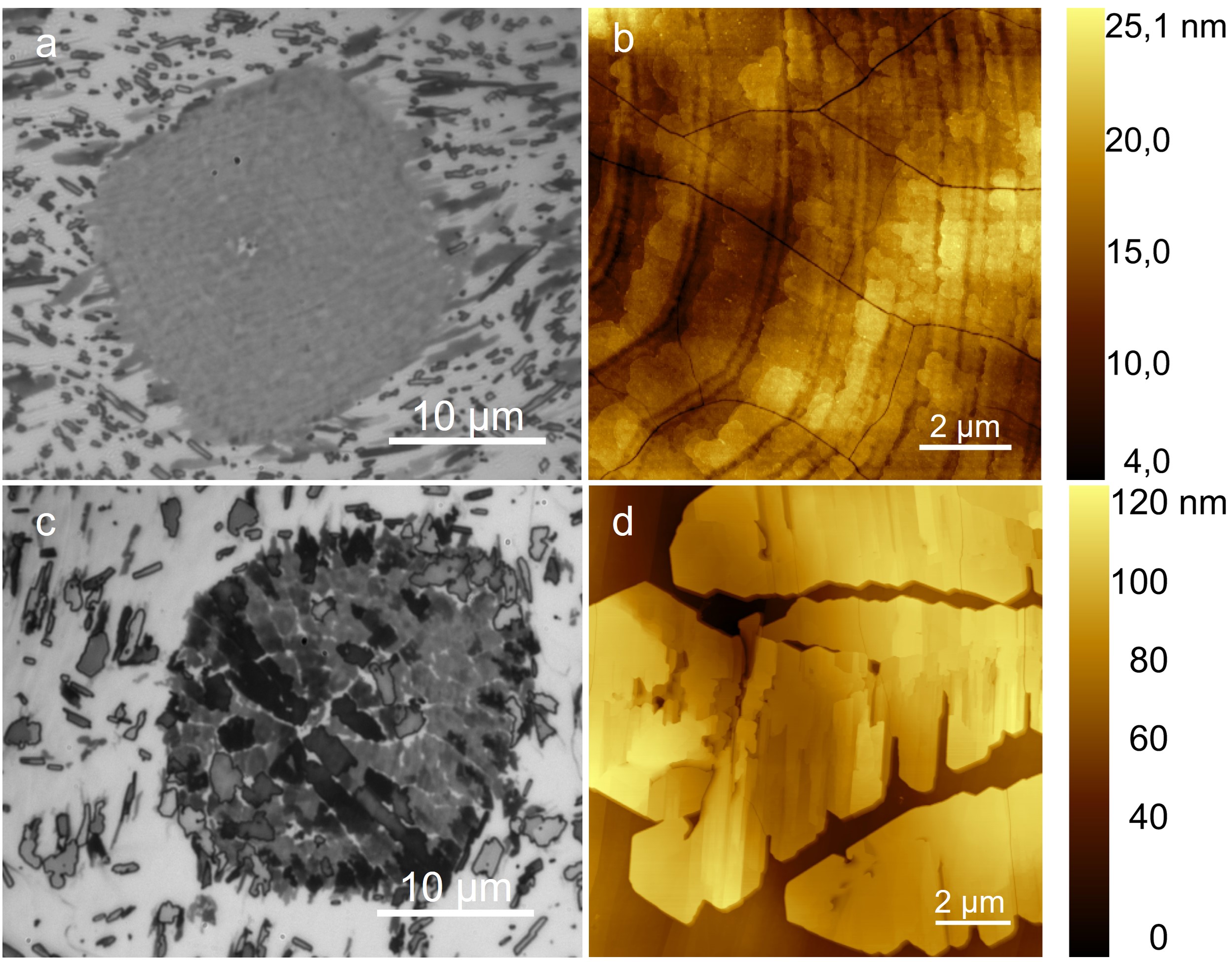}
    \caption{a and b: brightfield and AFM image of an orthorhombic platelet formed at the first high-rate heating process for 60 seconds. c and d: brightfield and AFM image of a platelet after performing the seconds high-rate heating showing the transform of the platelet to the separated bulk crystal. The step heights reveal the orthorhombic phase of rubrene crystalline structures.}
    \label{fig:evap_cons_heat_heat_bf_AFM}
\end{figure}

Based on the step heights on platelets which could be found all over the surface, these crystals belong to the orthorhombic phase with line and point defects. Although the orthorhombic phase is dominant when one heats the substrate at the second crystallization temperature which is around 170$^{\circ}$C, there is always the coexisting pathway of a transformation into the triclinic phase. The triclinic growth of rubrene, which is, as previously mentioned, a meta-stable phase, is particularly common for low-rate heating i.e., gradually heating the substrate from room temperature close to the first crystallization temperature around 140$^\circ$C. By gradually heating the substrate, the molecules have enough time to randomly attach to each other and form diffusion limited aggregates, which results in a dendritic, yet crystalline growth with a rough surface. In order to gain more insight into the competition between these phases, we now investigate the coverage dependency of their appearance, by means of a semi quantitative analysis \cite{Fielitz2016}.

One way to qualitatively analyze the coverage dependence is to inspect locations of large local coverage gradients, because a coverage dependency will be well visible as an inhomogeneous distribution of grown structures. We created such inhomogeneities by applying a shadow mask during vacuum evaporation, combined with an increased amount of rubrene (thickness: $\approx$50 nm). Figure \ref{fig:orth_tric_transi_bf}.a shows an overview brightfield image after high-rate heat treatment at 170$^{\circ}$C for one minute. The corresponding crystal structures were identified by comparing optical microscopy and AFM investigations (see also Section 3.1). A well defined transition line between orthorhombic and triclinic structures emerges at the shadowed boarder: in figure \ref{fig:orth_tric_transi_bf}.a orthorhombic ultra-flat and thin single crystals, indicated by arrows, are grown in the shadowed area where the rubrene coverage is extremely low due to low rubrene-on-HOPG diffusivity. But with an increase in rubrene coverage toward the border, orthorhombic platelets grow in the region marked by a dashed rectangle. They exhibit higher surface roughness which is even higher at escalated local rubrene coverage, leading to the growth of triclinic dendritic crystals. In the region right from the dashed rectangle, the growth is triclinic. Figure \ref{fig:orth_tric_transi_bf}.b shows a graph of the area covered by the different crystal species after one minute of high-rate heating as a function of rubrene layer thickness. In conclusion, for both, spin-coating and vacuum evaporation, higher coverage of rubrene favors the triclinic phase over orthorhombic one.

In view of the case for spin-coated samples, the question is whether crystal morphology and thickness is affected  by the number of heating steps in a similar fashion. Figure \ref{fig:evap_cons_heat_heat_bf_AFM}.a shoes a bright field image of a platelet prepared by evaporation of a 20 nm layer rubrene on HOPG and then high-rate heat treated for one minute at 170$^{\circ}$C. An AFM image of a related platelet in figure \ref{fig:evap_cons_heat_heat_bf_AFM}.a is shown in Figure \ref{fig:evap_cons_heat_heat_bf_AFM}.b. Figure \ref{fig:evap_cons_heat_heat_bf_AFM}.c shows the same sample after second, identical heat treatment with figure \ref{fig:evap_cons_heat_heat_bf_AFM}.d showing the related AFM image. These indicate that the thickness of the crystals increase by roughly 100 nm (see figure \ref{fig:evap_cons_heat_heat_bf_AFM}.b and \ref{fig:evap_cons_heat_heat_bf_AFM}.d) and the domains transform to single bulk crystalline packing upon a second high-rate heating step. The first heating step for one minute obviously does not provide enough energy to the system to let all the deposited rubrene contribute to the crystallization and the remaining material can then boost the thickness during the second heat treatment, a similar behavior as observed for spin-coating preparation (figure \ref{fig:cons_heat_bf_AFM}).

It was previously reported, for other substrates, that by heating the evaporated rubrene, platelets and spherulites start to grow preferentially in lateral size \cite{Fielitz2016} rather than in thickness. Upon further heating steps, we observe here that both the crystal morphology and thickness change for platelets and spherulites (see figure \ref{fig:evap_cons_heat_heat_bf_AFM}). This means there must be a reservoir (e.g. consisting of hardly visible few or monolayer rubrene) of rubrene which can reach the crystallites via transport across the surface, related to the process of Ostwald ripening \cite{Ribi2007}. It could be related to the better diffusivity on HOPG compared to other substrates.


\section{Conclusion}

Rubrene crystalline structures were prepared on bare HOPG using high-rate heating methods. The morphology of such molecular assemblies could be controlled by the amount and method of prior rubrene deposition, as well as by the duration, number and rate of heating treatments. Orthorhombic crystal growth is the result of low coverage while high coverage favours triclinic growth. The fluorescence intensities for these structures were investigated and we showed that the fluorescence response is a solid method to determine the quality and internal orientation of rubrene crystals. The orthorhombic crystal phase is much more resistant to photo-degradation, in contrast to the triclinic phase.

\section{Acknowledgement}

Funding by the Deutsche Forschungsgemeinschaft (DFG, German Research Foundation) within project number 441234705, SFB 1477 "Light-Matter Interactions at Interfaces" and project number 299150580, SFB 1270 "Electrically Active Implants" is acknowledged. We thank Prof. Stefan Lochbrunner and Tim Völzer for fruitful discussion and advice for the manuscript.

\bibliographystyle{unsrt}
\bibliography{Ref}
\end{document}